\documentclass[pre,twocolumn,showpacs,superscriptaddress]{revtex4-1}
\usepackage{graphicx}
\usepackage{amsmath}
\usepackage{multirow}
\usepackage{array}
\usepackage{chngcntr}

\makeatletter
\newcommand{\thickhline}{%
    \noalign {\ifnum 0=`}\fi \hrule height 1.2pt
    \futurelet \reserved@a \@xhline
}
\newcolumntype{"}{@{\hskip\tabcolsep\vrule width 1.2pt\hskip\tabcolsep}}
\makeatother

\begin{document}

\title{Chaos and unpredictability in evolution of cooperation in continuous time}
\author{Taekho You}
\affiliation{Department of Industrial and Management Engineering, Pohang University of Science and Technology, Pohang 37673, Korea}
\author{Minji Kwon}
\affiliation{Department of Industrial and Management Engineering, Pohang University of Science and Technology, Pohang 37673, Korea}
\author{Hang-Hyun Jo}
\email{hang-hyun.jo@apctp.org}
\affiliation{Asia Pacific Center for Theoretical Physics, Pohang 37673, Korea}
\affiliation{Department of Physics, Pohang University of Science and Technology, Pohang 37673, Korea}
\affiliation{Department of Computer Science, Aalto University, Espoo FI-00076,
Finland}
\author{Woo-Sung Jung}
\email{wsjung@postech.ac.kr}
\affiliation{Department of Industrial and Management Engineering, Pohang University of Science and Technology, Pohang 37673, Korea}
\affiliation{Asia Pacific Center for Theoretical Physics, Pohang 37673, Korea}
\affiliation{Department of Physics, Pohang University of Science and Technology, Pohang 37673, Korea}
\author{Seung Ki Baek}
\email{seungki@pknu.ac.kr}
\affiliation{Department of Physics, Pukyong National University, Busan 48513, Korea}

\begin{abstract}
Cooperators benefit others with paying costs. Evolution of cooperation
crucially depends on the cost-benefit ratio of cooperation, denoted as $c$.
In this work, we investigate the infinitely repeated prisoner's dilemma for
various values of $c$ with four of the representative memory-one strategies,
i.e., unconditional cooperation, unconditional defection, tit-for-tat, and
win-stay-lose-shift.  We consider replicator dynamics which deterministically
describes how the fraction of each strategy evolves over time in an
infinite-sized well-mixed population in the presence of implementation error and
mutation among the four strategies. Our finding is that this three-dimensional
continuous-time dynamics exhibits chaos through a bifurcation sequence similar
to that of a logistic map as $c$ varies. If mutation occurs with rate $\mu \ll
1$, the position of the bifurcation sequence on the $c$ axis is numerically
found to scale as $\mu^{0.1}$, and such sensitivity to $\mu$ suggests that
mutation may have non-perturbative effects on evolutionary paths. It
demonstrates how the microscopic randomness of the mutation process can be
amplified to macroscopic unpredictability by evolutionary dynamics.
\end{abstract}


\maketitle

\section{Introduction}

The notion of ``evolutionary progress'' has been debated ever since
Darwin~\cite{Gould1989Wonderful}, and
most evolutionary biologists dismiss the idea that evolution is directional
change toward the better~\cite{Shanahan2004Evolution}.
A counterexample to progressionism is self-extinction caused by individual
adaptation~\cite{Matsuda1994Runaway,*Gyllenberg2001Necessary,*Gyllenberg2002Evolutionary,*Rankin2005Adaptation},
which is directional change towards the worse by any measure.
For example, transgenic males of Japanese medaka fish {\it Oryzias latipes} have
larger body sizes than the wild-type counterparts and thus enjoy advantages in
mating, but they tend to decrease the population size because their
offspring have lower fecundity~\cite{Muir1999Possible}. Here, individual
interests contradict with the collective interest of the
population~\cite{Rankin2007Tragedy}, as is often modeled by
the prisoner's dilemma (PD) game. The PD game thus provides us with an
analytic model in which selection works in an undesirable direction.
However, one could even ask whether evolution is directional
after all, and progressionism will lose big ground if chaos proves
ubiquitous in evolution, i.e., marking natural history as unpredictable
alternations of progression and retrogression.
In fact, a recent study indicates that chaos becomes more likely as the
dimensionality of the phenotype space increases~\cite{Doebeli2014Chaos}, and
the possibility is greatly enhanced by discrete-time dynamics, especially
when one considers a wide range of
parameters~\cite{Vilone2011Chaos,*Galla2013Complex}.
For example, chaos in the iterated version of the
PD game has been reported among $10$ different
strategies under frequency-dependent selection in discrete
time~\cite{Nowak1993Chaos}. On the other hand,
chaos in low-dimensional continuous dynamics is a more challenging issue,
considering that chaos is impossible when the dimensionality of the
strategy space is less than three~\cite{Skyrms1992Chaos,*Strogatz2001Nonlinear}.
In this work, we report chaoticity in three-dimensional (3D)
continuous-time mutation-selection dynamics of the iterated PD game
with varying the cost-benefit ratio of cooperation.
By choosing the PD game as a microscopic foundation, we put our problem
into the context of conflict between individual and collective
interests in a biological population.
In particular, compared with the generic model in Ref.~\cite{Doebeli2014Chaos},
we will see that our approach directly relates unpredictability to mutation,
whereas self-extinction can be driven solely by selection,
so that it naturally incorporates another key argument of
non-progressionism, i.e., historical contingency of
mutation~\cite{Travisano1995Experimental,*Blount2008Historical,*Shah2015Contingency}.

This work is organized as follows: We explain the basic formulation of our
dynamical system in the next section. Section~\ref{sec:result} shows
numerical results, which will be discussed in Sec.~\ref{sec:discuss} from the
viewpoint of time reversal. We will then summarize this work in
Sec.~\ref{sec:summary}.

\section{Method}
\label{sec:method}

The PD game is a symmetric
two-person game defined by the following payoff table for the row player:
\begin{equation}
\left(
\begin{array}{c|cc}
   & C & D\\\hline
 C & b-c & -c\\
 D & b & 0\\
\end{array}
\right),
\label{eq:payoffs}
\end{equation}
where $C$ and $D$ denote two possible moves, i.e., cooperation and defection,
respectively. Note that the payoff table is parametrized by $b$ and $c$ which we
assume to satisfy $b>c>0$. If you choose to cooperate, it implies that it
benefits your coplayer by an amount of $b$ at the expense of your own cost $c$.
This is the reason that your payoff becomes $-c$
whereas the coplayer earns $+b$ when you cooperate and the coplayer defects
[Eq.~\eqref{eq:payoffs}].
Even if we begin with a group of cooperating individuals, a defecting trait will
invade and take over the population as soon as it comes into existence through
mutation, which is comparable to the scenario of
self-extinction~\cite{Matsuda1994Runaway,*Gyllenberg2001Necessary,*Gyllenberg2002Evolutionary,*Rankin2005Adaptation}.
The situation changes when the PD game is iterated between the pair of players
so that an individual may adopt a strategy according to which cooperation is
conditioned on the past interaction with the coplayer. A famous conditional
cooperator of this sort is tit-for-tat (TFT), which cooperates at the first
round and then copies the coplayer's previous move at each subsequent
round~\cite{Axelrod1984Evolution,*Nowak1992Tit,*Imhof2005Evolutionary}.
Another important strategy is
win-stay-lose-shift (WSLS), which attempts a different move from the previous
one if it did not earn a positive payoff~\cite{Kraines1989Pavlov,
*Nowak1993Strategy,*Posch1999Win,*Bladon2010Evolutionary,*Hilbe2017Memory}.
Both TFT and WSLS belong to a class of
memory-one strategies in the sense that they refer to only the previous round in
making decisions. These conditional cooperators achieve a high
level of cooperation based on reciprocity when the cost of cooperation is low
enough~\cite{Imhof2007Tit,*Hilbe2013Evolution,*Baek2016Comparing}.
Some animal societies seem to
have developed such behavior~\cite{Wilkinson1984Reciprocal, *Milinski1987Tit}.
To contrast those conditional strategies with unconditional ones, we denote
a strategy of unconditional cooperation (defection) as AllC (AllD) henceforth.
Although the unconditional strategies are memoryless, they are also
members of the memory-one strategy class with trivial dependence on the past
memory. These are the most extensively studied strategies, so the following
strategy set will be considered in this work: $\mathcal{S} \equiv \{ \mbox{AllC,
AllD, TFT, WSLS} \}$.

Let us consider the case when a player with strategy $i\in \mathcal{S}$ plays
the iterated PD game against the coplayer with strategy $j\in \mathcal{S}$. At
each round, they can occupy one of the following states: $(C,C)$, $(C,
D)$, $(D, C)$, and $(D, D)$. Suppose that the result is $(\alpha,\beta)$, or in
short $\alpha\beta$, at a current round ($\alpha, \beta \in \{C,D\}$).
Based on this result, each player's
strategy prescribes cooperation with certain probability, which is either zero
or one. The probability that the player (coplayer) with strategy $i$ ($j$)
cooperates at the next round is denoted by $q_{\alpha\beta}$
($r_{\beta\alpha}$). For example, for a player with AllC, $q_{\alpha\beta}=1$
for all states of $\alpha\beta$, while for a player with TFT,
$q_{_{CC}}=q_{_{DC}}=1$ and $q_{_{CD}}=q_{_{DD}}=0$ (for details, see
Appendix~\ref{sect:payoff}). The values of $r_{\beta\alpha}$ can be similarly
assigned. In the presence of implementation error with probability $\epsilon \in
(0,1)$, the probabilities are effectively changed to $q'_{\alpha\beta} =
(1-\epsilon) q_{\alpha\beta} + \epsilon (1 - q_{\alpha\beta})$ and
$r'_{\beta\alpha} = (1-\epsilon) r_{\beta\alpha} + \epsilon (1 -
r_{\beta\alpha})$, respectively. By defining $\bar{q}'_{\alpha\beta} \equiv
1-q'_{\alpha\beta}$ and $\bar{r}'_{\beta\alpha} \equiv 1-r'_{\beta\alpha}$, the
transition matrix between the current and next states is written as
\begin{equation}
M =
\begin{bmatrix}
q'_{_{CC}} r'_{_{CC}} &
q'_{_{CD}} r'_{_{DC}} &
q'_{_{DC}} r'_{_{CD}} &
q'_{_{DD}} r'_{_{DD}} \\
q'_{_{CC}} \bar{r}'_{_{CC}} &
q'_{_{CD}} \bar{r}'_{_{DC}} &
q'_{_{DC}} \bar{r}'_{_{CD}} &
q'_{_{DD}} \bar{r}'_{_{DD}} \\
\bar{q}'_{_{CC}} r'_{_{CC}} &
\bar{q}'_{_{CD}} r'_{_{DC}} &
\bar{q}'_{_{DC}} r'_{_{CD}} &
\bar{q}'_{_{DD}} r'_{_{DD}} \\
\bar{q}'_{_{CC}} \bar{r}'_{_{CC}} &
\bar{q}'_{_{CD}} \bar{r}'_{_{DC}} &
\bar{q}'_{_{DC}} \bar{r}'_{_{CD}} &
\bar{q}'_{_{DD}} \bar{r}'_{_{DD}}
\end{bmatrix},
\end{equation}
which admits a unique right eigenvector $\vec{v} = \left(v_{_{CC}}, v_{_{CD}},
v_{_{DC}}, v_{_{DD}}\right)$ with eigenvalue one according to the
Perron-Frobenius theorem. We normalize $\vec{v}$ by requiring $v_{_{CC}} +
v_{_{CD}} + v_{_{DC}} + v_{_{DD}} = 1$ to take it as stationary probability
distribution over the four states. Note that the existence of error with
$\epsilon > 0$ makes the initial state irrelevant in the long run. This
procedure is applied to every pair of strategies $i$ and $j$ in $\mathcal{S}$.
By taking the inner product between the resulting $\vec{v}$ with a payoff vector
$(b-c, -c, b, 0)$, we calculate the long-term payoff $a_{ij}$ that strategy $i$
obtains against $j$ (see Appendix~\ref{sect:payoff} for details).

Let $x_i(t)$ denote the fraction of strategy $i \in \mathcal{S}$ at time $t$.
If the population is infinitely large, $x_i$ can be regarded as a real variable.
The payoff of strategy $i$, denoted by $p_i(t)$, is given as $p_i(t) = \sum_j
a_{ij} x_j(t)$ if the population is well-mixed. The population average payoff
will thus be $\left< p \right>(t) \equiv \sum_i p_i(t) x_i(t) = \sum_{ij} a_{ij}
x_i(t) x_j(t)$. We describe the mutation-selection dynamics of the population
using the replicator dynamics
(RD)~\cite{Weibull1995Evolutionary,Hofbauer1998Evolutionary} as follows:
\begin{equation}
\frac{dx_i}{dt} = f_i \equiv (1-\mu) p_i x_i - \left< p \right> x_i +
\frac{\mu}{|\mathcal{S}|-1} \sum_{j \neq i} p_j x_j,
\label{eq:rd}
\end{equation}
where $\mu$ denotes the rate of mutation.
We have only three degrees
of freedom because $|\mathcal{S}|=4$ and $\sum_{i\in\mathcal{S}} x_i(t) = 1$ all
the time. The system described by Eq.~\eqref{eq:rd} is therefore understood as a
deterministic dynamical system inside a 3D region, defined as
$\Omega = \{ \mathbf{x} \equiv (x_{_{\rm AllD}}, x_{_{\rm TFT}}, x_{_{\rm
WSLS}}) | x_{_{\rm AllD}} \ge 0$, $x_{_{\rm TFT}} \ge 0$, $x_{_{\rm WSLS}} \ge
0$, and $x_{_{\rm AllD}} + x_{_{\rm TFT}} + x_{_{\rm WSLS}} \le 1 \}$. Note that
the time scale governing the RD has been assumed to be much larger than that for
calculating the ``long-term'' payoffs. Our main finding is that the dynamics
of Eq.~\eqref{eq:rd} exhibits chaos in this $\Omega$ when $\mu > 0$. Our system
is minimal to be chaotic, and this minimal system helps us understand the
role of mutation in analytic terms.

\begin{figure}
    \includegraphics[width=\columnwidth]{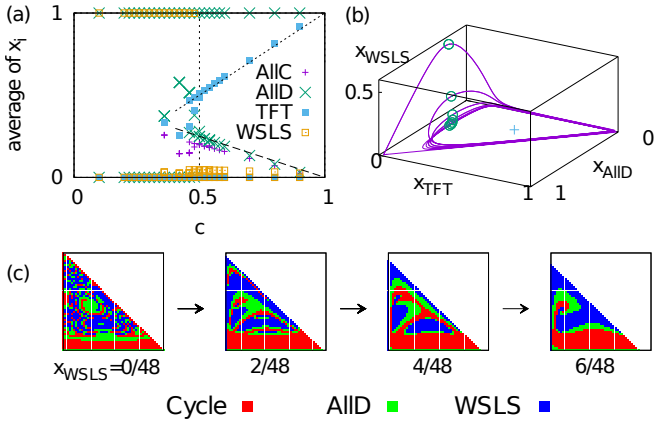}
    \caption{(Color online) (a) Fractions of strategies averaged over long time
    [Eq.~(\ref{eq:avgx})], started from various initial conditions (see the main
    text). The vertical dotted line indicates $c=\frac{1}{2}$, and the two
    oblique dotted lines represent $x_{_{\rm TFT}} \approx c$ and $x_{_{\rm
    AllD}} \approx \frac{1-c}{2}$, respectively.
    (b) Attractor for $c=0.46$ as a stable limit cycle, which rotates
    counterclockwise if projected onto the ($x_{_{\rm TFT}},x_{_{\rm AllD}}$)
    plane. The small circles represent local maxima of $x_{_{\rm WSLS}}$ along
    the trajectory, and the cross indicates a mutation-free fixed point (FP)
    written as $\mathbf{x}^*_0 = \left( \frac{1-2\epsilon-c}{2-4\epsilon},
    \frac{c}{1-2\epsilon}, 0 \right)$.
    (c) Basins of attraction for $c=0.46$, where the horizontal and vertical
    coordinates indicate $x_{_{\rm AllD}}$ and $x_{_{\rm TFT}}$,
    respectively. These are cross-sectional views at different values of
    $x_{_{\rm WSLS}}$ from $\frac{0}{48}$ to $\frac{6}{48}$. The basin of
    attraction for a cycle as shown in (b) is denoted in red, AllD
    FP with $\mathbf{x}^*=(1,0,0)$ in green, and WSLS FP
    with $\mathbf{x}^*=(0,0,1)$ in blue.}
    \label{fig:avgx}
\end{figure}

\section{Result}
\label{sec:result}

Numerically, we integrate this set of ordinary differential equations for every
$i \in \mathcal{S}$ by means of the fourth-order Runge-Kutta method, with
enforcing the normalization condition $\sum_{i\in\mathcal{S}} x_i(t) = 1$ at any
time $t$. We set $b = 1$ without loss of generality, and take $\epsilon =
10^{-2}$ as the probability of error.
We choose $\Delta t = 10^{-2}$ as the incremental time interval for
numerical integration throughout this work. Note that the RD depends on the
initial condition at $t=0$ due to its deterministic nature. It is therefore
important to probe many different initial conditions to obtain the full picture.
In Fig.~\ref{fig:avgx}(a), we assume $\mu=10^{-4}$ and plot the time average of $x_i$,
\begin{equation}
    \bar{x}_i \equiv \frac{1}{t_2 - t_1} \int_{t_1}^{t_2} x_i(t) dt,
\label{eq:avgx}
\end{equation}
which removes transient dynamics for $0 \leq t < t_1$ with $t_1 \gg O(1)$ and
considers the history up to $t_2 > t_1$. For each value of $c$, we have checked
roughly $2\times 10^4$ different initial conditions for $\mathbf{x} = (x_{_{\rm
AllD}}, x_{_{\rm TFT}}, x_{_{\rm WSLS}})$ in $\mathcal{I}_\delta \equiv \{ (0,0,0)$,
$(0, 0, \delta)$, $(0, 0, 2\delta), \ldots, (1, 0, 0) \}$ with $\delta =
\frac{1}{48}$. If we take $c=0.2$, for example, we find from
Fig.~\ref{fig:avgx}(a) that the system has two different attractors,
characterized by $\mathbf{x}^* \approx (1, 0, 0)$ and $\mathbf{x}^* \approx (0,
0, 1)$, respectively, depending on the initial condition. The latter fixed point
(FP) disappears as $c$ roughly exceeds $\frac{1}{2}$, and this is readily
explained by the FP analysis: When $\mu=0$, Eq.~\eqref{eq:rd} has 11 different
FP's, one of which is $\mathbf{x}^*_0 = (0, 0, 1)$. Its eigenvalues are
all negative when $c$ lies below a certain threshold $c_{\rm th} =
\frac{(1-2\epsilon )^2} {2-4\epsilon +4\epsilon^2} \approx \frac{1}{2} -
\epsilon$, whereas one of them becomes positive for $c > c_{\rm th}$ (see Appendix~\ref{sect:mu0} for details). In terms of evolutionary biology, we can also say that
the WSLS-dominant FP is an evolutionarily stable state (ESS) for $c <
c_{\rm th}$, noting that an ESS constitutes an asymptotically stable FP
in RD~\cite{Weibull1995Evolutionary}. Figure~\ref{fig:avgx}(a) also shows that
the distribution of attractors changes qualitatively as $c$ increases. Roughly
speaking, if $c \gtrsim 0.4$, we can observe another attractor whose time
average is approximated as $\bar{\mathbf{x}} = (\bar{x}_{_{\rm AllD}},
\bar{x}_{_{\rm TFT}}, \bar{x}_{_{\rm WSLS}}) \approx \left( \frac{1-c}{2}, c,
\eta \right)$ with a small positive number $\eta$. This attractor actually
corresponds to a cycle, and Fig.~\ref{fig:avgx}(b) shows an example for
$c=0.46$. By a cycle, we do not only mean an ordinary limit cycle of finite
periodicity but also a quasi-periodic or strange attractor of an infinite
period. Figure~\ref{fig:avgx}(c) provides tomographic views of the basins of
attraction with different values of $x_{_{\rm WSLS}}(t=0)$.
The basins have very complicated boundaries on the $(x_{_{\rm AllD}}, x_{_{\rm TFT}})$ plane and gradually merge as we increase $x_{_{\rm WSLS}}$ in the initial condition.
Therefore, although WSLS can stabilize cooperation for
$c < c_{\rm th}$, it depends on initial conditions, and one cannot easily choose
a correct one if beginning with low $x_{_{\rm WSLS}}$.

\begin{figure}
    \includegraphics[width=\columnwidth]{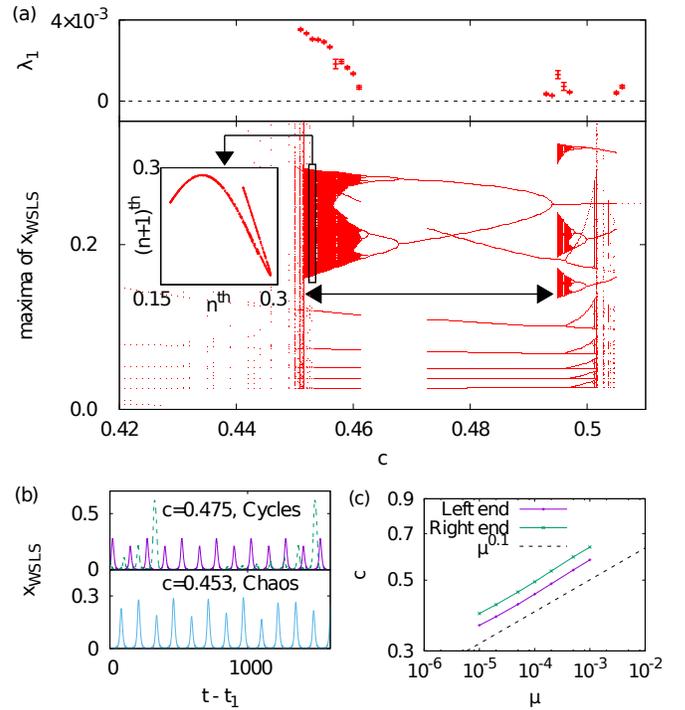}
    \caption{(Color online) (a) Largest Lyapunov exponent $\lambda_1$ when it is
    positive, and local maxima of $x_{_{\rm WSLS}}(t)$ [see the circles in
    Fig.~\ref{fig:avgx}(b)] at various values of $c$. We check
    many different cycles to estimate the average and standard error of
    $\lambda_1$. The big horizontal arrow shows the size of the main
    bifurcation structure, and the long vertical rectangle indicates $c=0.453$,
    at which chaos fully develops. Inset: return map between the $n$th and
    $(n+1)$th maxima of $x_{\rm WSLS}$ at $c=0.453$.
    (b) Time series of $x_{\rm WSLS}$ at $c=0.475$ (upper) and $c=0.453$
    (lower), after removing transient dynamics for $t_1 \gg O(1)$. The two lines
    of the upper panel are obtained with different initial conditions.
    (c) Left and right ends of the main
    bifurcation structure [see the horizontal arrow in (a)]
    as a function of $\mu$ on the log-log scale. We have
    additionally plotted $\mu^{0.1}$ for comparison.}
\label{fig:bif}
\end{figure}

The complicated geometry of the basins of attraction suggests
the possibility of chaos~\cite{McDonald1985Fractal}. To quantify
chaoticity, we calculate Lyapunov exponents with
LyapOde4~\cite{Bryant2009LyapOde}, which is based on the QR decomposition
method~\cite{Eckmann1985Ergodic,*Abarbanel1991Lyapunov}.
In Fig.~\ref{fig:bif}(a), we plot the largest Lyapunov exponent $\lambda_1$ when
it is estimated as positive (see Appendix~\ref{sect:lyap} for details). The result
supports the existence of chaos. Furthermore, we can explicitly
construct a bifurcation diagram [Fig.~\ref{fig:bif}(a)] by plotting local maxima
of $x_{_{\rm WSLS}}$ [the circles in Fig.~\ref{fig:avgx}(b)] at various values
of $c$. The bifurcation structure is reminiscent of that of the logistic map,
and the similarity can be made more precise by plotting a return map between
the $n$th and $(n+1)$th maxima when the system is chaotic [inset of
Fig.~\ref{fig:bif}(a)].
We also note that Fig.~\ref{fig:bif}(a) shows many other lines than the
bifurcation diagram because two different cycles can coexist at the same $c$
depending on their initial conditions. For example,
the upper panel of Fig.~\ref{fig:bif}(b) depicts two time series of $x_{_{\rm
WSLS}}$ at $c=0.475$: the solid line shows two alternating peaks, whereas the
dashed line is characterized by gradually growing peaks followed by a quiet
stage. As $c$ decreases, the former cycle undergoes period doubling to chaos
[the lower panel in Fig.~\ref{fig:bif}(b)],
but the latter one still remains stable with finite periodicity.
Although unpredictable, the dynamics exhibits a variety of correlations that we
can make use of. For example, the maximum of $x_{_{\rm WSLS}}$ provides
a precursor for that of $x_{_{\rm AllD}}$ because
it is correlated with how close the trajectory gets to the state of $x_{_{\rm
AllD}}=1$ before turning toward the state of $x_{_{\rm TFT}}=1$
[see Fig.~\ref{fig:avgx}(b)].

The position of the main bifurcation structure is extremely
sensitive to the variation of $\mu$: Fig.~\ref{fig:bif}(c) shows the left and
right ends of the main bifurcation structure, represented by the horizontal
arrow in Fig.~\ref{fig:bif}(a). According to our numerical observation, they are
roughly proportional to $\mu^{0.1}$. If the power-law behavior holds for
$\mu \rightarrow 0$, the bifurcation structure will shrink and eventually
disappear as mutation occurs less and less frequently. The exponent is
unexpectedly small, and this sensitivity to $\mu$ is consistent with
the following observation:
Suppose that we have a FP $\mathbf{x}^*_0$ when mutation
is absent, i.e., $\mu=0$. The corresponding FP $\mathbf{x}^*$ in the
presence of small $\mu > 0$ cannot be obtained directly but approximated through
a perturbative calculation~\cite{Baek2017Duality}. By denoting
$\mathbf{f}(\mathbf{x}) \equiv (f_{_{\rm AllD}}(\mathbf{x}), f_{_{\rm
TFT}}(\mathbf{x}), f_{_{\rm WSLS}}(\mathbf{x}))$, we obtain
\begin{equation}
    \mathbf{x}^* \approx \mathbf{x}^*_0 - \mathbf{J}(\mathbf{x}^*_0)^{-1}\cdot \mathbf{f}(\mathbf{x}^*_0),
\label{eq:newton2}
\end{equation}
if the Jacobian matrix $\mathbf{J}=\left\{ \frac{\partial f_i}{\partial x_j}
\right\}$ has an inverse at $\mathbf{x}=\mathbf{x}^*_0$ (see Appendix~\ref{sect:pert} for
details). The last term on the right-hand side of Eq.~\eqref{eq:newton2}
describes displacement of the FP due to mutation and it is expected to
be of an order of $\mu$ in this perturbative approach. However, this is not the
case if we look at a mutation-free FP $\mathbf{x}^*_0 = \left(
\frac{1-2\epsilon-c}{2-4\epsilon}, \frac{c}{1-2\epsilon}, 0 \right)$, which is
close to $\bar{\mathbf{x}} \approx \left( \frac{1-c}{2}, c, \eta \right)$, the
time average of the cycle in Fig.~\ref{fig:avgx}(b):
We rather find that $\frac{\partial f_{\rm WSLS}}{\partial x_i}$ is only of an
order of $\mu$ for every $i \in \mathcal{S}$, so that the naive perturbative
analysis of Eq.~\eqref{eq:newton2} yields displacement of $O(1)$. Note that it
does not vanish in the limit of $\mu \rightarrow 0$, contrary to our
expectation. The failure in the perturbative estimate of $\mathbf{x}^*$
nevertheless suggests that mutation can greatly affect the FP structure near the
limit cycle.

\section{Discussion}
\label{sec:discuss}

Whenever an ESS exists, it is an asymptotically stable FP in the
mutation-free RD, i.e., Eq.~\eqref{eq:rd} with $\mu=0$, and a local Lyapunov
function can be constructed in the vicinity of this
FP~\cite{Weibull1995Evolutionary, Hofbauer1998Evolutionary}. However, RD may
have different types of attractors such as a limit cycle and a strange
attractor, and we then have no systematic way to regard the dynamic evolution as
optimization of a certain target function. In this context, it is worth noting
that RD has emergent symmetry~\cite{Baek2017Duality}: Let us denote
\begin{eqnarray}
    dx_{_{\rm AllC}}/dt &\equiv& f_{_{\rm AllC}} (x_{_{\rm AllC}}, x_{_{\rm AllD}})
    \label{eq:allc}\\
    dx_{_{\rm AllD}}/dt &\equiv& f_{_{\rm AllD}} (x_{_{\rm AllC}}, x_{_{\rm AllD}})
    \label{eq:alld}
\end{eqnarray}
with setting $x_{_{\rm WSLS}} = \mu = 0$. Then, the following equality
holds:
\begin{equation}
    f_{_{\rm AllC}} (x_{_{\rm AllC}}, x_{_{\rm AllD}}) + f_{_{\rm AllD}}
    (x_{_{\rm AllD}}, x_{_{\rm AllC}}) = 0.
    \label{eq:reversal}
\end{equation}
Therefore, if we redefine $X_{_{\rm AllC}} \equiv x_{_{\rm AllD}}$, $X_{_{\rm AllD}} \equiv x_{_{\rm AllC}}$, and $\tau \equiv -t$, we find from Eq.~(\ref{eq:reversal}) that
\begin{eqnarray}
    dX_{_{\rm AllC}}/d\tau &=& f_{_{\rm AllC}} (X_{_{\rm AllC}}, X_{_{\rm AllD}})\\
    dX_{_{\rm AllD}}/d\tau &=& f_{_{\rm AllD}} (X_{_{\rm AllC}}, X_{_{\rm AllD}}),
\end{eqnarray}
recovering the original dynamics in Eqs.~\eqref{eq:allc} and \eqref{eq:alld}
(see Appendix~\ref{sect:symm} for more detailed derivation). In
other words, if both $x_{_{\rm WSLS}}$ and $\mu$ are strictly zero, the dynamics
is dual to itself under time reversal and exchange between AllC and AllD. The
duality implies that our two-dimensional dynamics cannot have a global Lyapunov
function: If a function $V(x_{_{\rm AllC}}=x_1, x_{_{\rm AllD}}=x_2)$ has a
positive time derivative, $V(x_2, x_1)$ must be a decreasing
function of time. Loosely speaking, therefore, if WSLS was absent,
it would be impossible to define the arrow of time for the whole system based
solely on selection.
We have two ways for directionality at this point: One is to incorporate
$x_{_{\rm WSLS}} > 0$ into dynamics to let the
system converge to the corresponding ESS through \emph{selection}. The
other is to keep $x_{_{\rm WSLS}}$ at zero but introduce $\mu>0$ and redefine
the arrow of time in terms of \emph{mutation}. For example, the FP
$\mathbf{x}^*$, modified from $\mathbf{x}^*_0 = \left(
\frac{1-2\epsilon-c}{2-4\epsilon}, \frac{c}{1-2\epsilon}, 0 \right)$ by
mutation, has eigenvalues with a negative real part proportional to
$\mu$~\cite{Baek2017Duality}. It implies that mutation mixes up the populations
of different strategies so that the system always ends up with the same
polymorphic state.
In either case, the initial condition eventually becomes irrelevant.
Our finding implies that if both $x_{_{\rm WSLS}}$ and $\mu$ are turned on,
the symmetry under time reversal and AllC-AllD exchange can break in a more
nontrivial way.
In particular, we have seen that mutation, induced by the inherent microscopic
randomness of the environment, produces unpredictability on evolutionary scales.
One might say that this is not unexpected because mutation {\it per se} is a
random event which makes the evolutionary path deviate from the one determined
by the initial condition. However, we are dealing with mutation in a fully
deterministic manner in Eq.~\eqref{eq:rd}, and the effect of mutation is not
just small perturbation added to a deterministic path, but the FP
structure itself seems to experience non-perturbative changes.
The chaotic dynamics is directional in a rather subtle sense that the path
reveals more and more information of the initial condition as time goes by.
Differently from the above cases, this additional information is not a
function of the current state exclusively, but something evaluated with
reference to history~\cite{Wackerbauer1994Comparative}. Historical contingency
should be understood in this respect, considering that not all history-dependent
systems are chaotic.

\section{Summary}
\label{sec:summary}

In summary, we have presented numerical evidence that chaos exists in the
3D continuous-time RD with mutation among the four representative
strategies of the PD game, i.e., AllC, AllD, TFT, and WSLS. Even if one
considers a more general strategy space, it will contain these four in most
cases, and our findings will remain valid in the corresponding subspace.
The model, originally motivated by selection against the better for the
population,
provides a simple analytic picture for non-progressionism, showing
various facets of mutation: It generates unpredictability with exerting
non-perturbative effects on evolutionary paths and breaks symmetry of a
subsystem under time reversal.
The strategic interaction considered here is one of the most intensively
studied subjects in evolutionary game theory, so our finding suggests that chaos
can be more prevalent than previously known.

\begin{acknowledgments}
H.-H.J. acknowledges financial support by Basic Science Research Program through
the National Research Foundation of Korea (NRF) grant funded by the Ministry of
Education (Grant No. 2015R1D1A1A01058958).
W.-S.J. was supported by Basic Science Research Program through the National
Research Foundation of Korea (NRF) funded by the Ministry of
Education (Grant No. NRF-2016R1D1A1B03932590).
S.K.B. was supported by Basic Science Research Program through the National
Research Foundation of Korea (NRF) funded by the Ministry of Science, ICT and
Future Planning (Grant No. NRF-2017R1A1A1A05001482).
\end{acknowledgments}

\appendix
\counterwithin{figure}{section}
\counterwithin{table}{section}

\section{Derivation of long-term payoff $a_{ij}$}
\label{sect:payoff}


As mentioned in the main text, we denote by $q_{\alpha\beta}$ ($r_{\beta\alpha}$) the probability that the player (coplayer) with strategy $i$ ($j$) cooperates at the next round, given that the state of player's and coplayer's moves is $(\alpha,\beta)$ at the current round. Here $\alpha,\beta\in\{C,D\}$. We summarize the values of $q_{\alpha\beta}$ for each strategy $i\in \mathcal{S} = \{ \mbox{AllC, AllD, TFT, WSLS} \}$ as follows:
\begin{center}
\begin{tabular}{c"c|c|c|c}
\thickhline
$(\alpha,\beta)$ &AllC & AllD & TFT & WSLS\\ \thickhline
	$(C,C)$ & 1 & 0 & 1 & 1\\ \hline
	$(C,D)$ & 1 & 0 & 0 & 0\\ \hline
	$(D,C)$ & 1 & 0 & 1 & 0\\ \hline
	$(D,D)$ & 1 & 0 & 0 & 1\\
\thickhline
\end{tabular}
\end{center}
The values of $r_{\beta\alpha}$ for the coplayer with strategy $j$ can be similarly assigned.


In the presence of implementation error with probability $\epsilon \in (0,1)$, the probabilities of cooperating are effectively changed to $q'_{\alpha\beta} = (1-\epsilon) q_{\alpha\beta} + \epsilon (1 - q_{\alpha\beta})$ and $r'_{\beta\alpha} = (1-\epsilon) r_{\beta\alpha} + \epsilon (1 - r_{\beta\alpha})$, respectively. By defining $\bar{q}'_{\alpha\beta} \equiv 1-q'_{\alpha\beta}$ and $\bar{r}'_{\beta\alpha} \equiv 1-r'_{\beta\alpha}$, the transition matrix between the current and next states is written as
\begin{equation}
M =
\begin{bmatrix}
q'_{_{CC}} r'_{_{CC}} &
q'_{_{CD}} r'_{_{DC}} &
q'_{_{DC}} r'_{_{CD}} &
q'_{_{DD}} r'_{_{DD}} \\
q'_{_{CC}} \bar{r}'_{_{CC}} &
q'_{_{CD}} \bar{r}'_{_{DC}} &
q'_{_{DC}} \bar{r}'_{_{CD}} &
q'_{_{DD}} \bar{r}'_{_{DD}} \\
\bar{q}'_{_{CC}} r'_{_{CC}} &
\bar{q}'_{_{CD}} r'_{_{DC}} &
\bar{q}'_{_{DC}} r'_{_{CD}} &
\bar{q}'_{_{DD}} r'_{_{DD}} \\
\bar{q}'_{_{CC}} \bar{r}'_{_{CC}} &
\bar{q}'_{_{CD}} \bar{r}'_{_{DC}} &
\bar{q}'_{_{DC}} \bar{r}'_{_{CD}} &
\bar{q}'_{_{DD}} \bar{r}'_{_{DD}}
\end{bmatrix}.
\end{equation}
Then, from the equation of $M\vec{v} = \vec{v}$, we obtain the stationary solution as $\vec{v}=\left(v_{_{CC}}, v_{_{CD}}, v_{_{DC}}, v_{_{DD}}\right)$, which is a unique right eigenvector of the matrix $M$. Here we impose the normalization for $\vec{v}$ using $v_{_{CC}}+ v_{_{CD}}+ v_{_{DC}}+ v_{_{DD}}=1$. Then $v_{\alpha\beta}$ can be interpreted as the stationary probability of finding the state $(\alpha,\beta)$ when players with strategies $i$ and $j$ play the game. The results of $\vec{v}$ for all combinations of $i$ and $j$ are summarized in Table~\ref{ta:stationary}. Once $\vec{v}$ is obtained, we calculate the long-term payoff $a_{ij}$ that the strategy $i$ obtains against $j$, by taking the inner product between $\vec{v}$ and the payoff vector such that
\begin{equation}
    a_{ij}=\vec{v}\cdot (b-c,-c,b,0).
\end{equation}
The long-term payoffs $a_{ij}$ for all possible pairs of $i$ and $j$ are presented in Table~\ref{ta:longtermpayoff}. Note that with $\epsilon>0$, initial states are irrelevant in the long run.

\begin{table*}[!ht]
    \caption{Stationary probability distributions $v_{\alpha\beta}$ of finding the state $(\alpha,\beta)$ when strategy $i$ plays against $j$, where $(\alpha,\beta)$ can be one of the four states: $(C,C)$, $(C,D)$, $(D,C)$, and $(D,D)$.}
\begin{tabular}{c|c"p{0.2\textwidth}|p{0.2\textwidth}|p{0.2\textwidth}|p{0.2\textwidth}}
\thickhline
	$i$ & $j$ &\multicolumn{1}{c|}{$v_{_{CC}}$} &\multicolumn{1}{c|}{$v_{_{CD}}$} &\multicolumn{1}{c|}{$v_{_{DC}}$} &\multicolumn{1}{c}{$v_{_{DD}}$}\\ \thickhline
\multirow{4}{*}{AllC}
	& AllC & $(1-\epsilon)^2$ & $\epsilon(1-\epsilon)$ & $\epsilon(1-\epsilon)$ & $\epsilon^2$ \\ \cline{2-6}
	& AllD & $\epsilon(1-\epsilon)$ & $(1-\epsilon)^2$ & $\epsilon^2$ & $\epsilon(1-\epsilon)$ \\ \cline{2-6}
	& TFT & $1-3\epsilon+4\epsilon^2-2\epsilon^3$ & $2\epsilon(1-\epsilon)^2$ & $\epsilon(1-2\epsilon-2\epsilon^2)$ & $2\epsilon^2(1-\epsilon)$\\ \cline{2-6}
	& WSLS & $(1-\epsilon)/2$ & $(1-\epsilon)/2$ & $\epsilon/2$ & $\epsilon/2$ \\ \thickhline
\multirow{4}{*}{\rm AllD}
	& AllC & $\epsilon(1-\epsilon)$ & $\epsilon^2$ & $(1-\epsilon)^2$ & $\epsilon(1-\epsilon)$ \\ \cline{2-6}
	& AllD & $\epsilon^2$ & $\epsilon(1-\epsilon)$ & $\epsilon(1-\epsilon)$ & $(1-\epsilon)^2$ \\ \cline{2-6}
	& TFT & $2\epsilon^2(1-\epsilon)$ & $\epsilon(1-2\epsilon-2\epsilon^2)$ & $2\epsilon(1-\epsilon)^2$ & $1-3\epsilon+4\epsilon^2-2\epsilon^3$ \\ \cline{2-6}
	& WSLS & $\epsilon/2$ & $\epsilon/2$ & $(1-\epsilon)/2$ & $(1-\epsilon)/2$ \\ \thickhline
\multirow{4}{*}{\rm TFT}
	& AllC & $1-3\epsilon+4\epsilon^2-2\epsilon^3$ & $\epsilon(1-2\epsilon+2\epsilon^2)$ & $2\epsilon(1-\epsilon)^2$ & $2\epsilon^2(1-\epsilon)$ \\ \cline{2-6}
	& AllD & $2\epsilon^2(1-\epsilon)$ & $2\epsilon(1-\epsilon)^2$ & $\epsilon(1-2\epsilon+2\epsilon^2)$ & $1-3\epsilon+4\epsilon^2-2\epsilon^3$ \\ \cline{2-6}
	& TFT & $1/4$ & $1/4$ & $1/4$ & $1/4$ \\ \cline{2-6}
	& WSLS & $1/4$ & $1/4$ & $1/4$ & $1/4$ \\ \thickhline
\multirow{4}{*}{\rm WSLS}
	& AllC & $(1-\epsilon)/2$ & $\epsilon/2$ & $(1-\epsilon)/2$ & $\epsilon/2$ \\ \cline{2-6}
	& AllD & $\epsilon/2$ & $(1-\epsilon)/2$ & $\epsilon/2$ & $(1-\epsilon)/2$ \\ \cline{2-6}
	& TFT & $1/4$ & $1/4$ & $1/4$ & $1/4$ \\ \cline{2-6}
	& WSLS & $1-4\epsilon+7\epsilon^2-4\epsilon^3$ & $\epsilon(1-\epsilon)$ & $\epsilon(1-\epsilon)$ & $\epsilon(2-5\epsilon+4\epsilon^2)$ \\
\thickhline
\end{tabular}
\label{ta:stationary}
\end{table*}


\begin{table}[!ht]
    \caption{Long-term payoff $a_{ij}$ that strategy $i$ (leftmost column) obtains against $j$ (top row).}
\begin{tabular}{c"p{0.2\columnwidth}|p{0.2\columnwidth}|p{0.2\columnwidth}|p{0.2\columnwidth}}
\thickhline
    & \multicolumn{1}{c|}{AllC} & \multicolumn{1}{c|}{AllD}  & \multicolumn{1}{c|}{TFT}  & \multicolumn{1}{c}{WSLS}  \\ \thickhline
	AllC & $(b-c)(1-\epsilon)$ & $b\epsilon-c(1-\epsilon)$ & $b(1-2\epsilon+2\epsilon^2)-c(1-\epsilon)$ & ${b\over2}-c(1-\epsilon)$ \\ \hline
	AllD & $b(1-\epsilon)-c\epsilon$ & $(b-c)\epsilon$ & $(2b(1-\epsilon)-c)\epsilon$ & ${1\over2}(b-2c\epsilon)$ \\ \hline
	TFT & $b(1-\epsilon)-c(1-2\epsilon-2\epsilon^2)$ & $(b-2c(1-\epsilon))\epsilon$ & ${b-c\over2}$ & ${b-c\over2}$ \\ \hline
	WSLS & $b(1-\epsilon)-{c\over2}$ & $b\epsilon-{c\over2}$ & ${b-c\over2}$ & $(b-c)(1-3\epsilon+6\epsilon^2-4\epsilon^3)$\\ \thickhline
\end{tabular}
\label{ta:longtermpayoff}
\end{table}

\section{FP analysis of the replicator dynamics when $\mu = 0$}
\label{sect:mu0}

In order to better understand the RD in Eq. (3) in the
main text, we first analyze the mutation-free case, i.e., when $\mu=0$.
Let us define
$\mathbf{x} \equiv (x_{_{\rm AllD}}, x_{_{\rm TFT}}, x_{_{\rm WSLS}})$,
where $x_i$ denotes the fraction of strategy $i$.
From the set of equations
\begin{equation}
    0=\frac{dx_i}{dt}=f_i(\mu=0)=p_ix_i - \left< p\right>x_i
\end{equation}
for all $i\in \mathcal{S}$, we analytically find $11$ different FP's
$\mathbf{x}_0^*$ as follows:
\begin{equation}
\mathbf{x}_0^* =
\begin{cases}
(1,0,0)\\
(0,1,0)\\
(0,0,1)\\
(0,0,0)\\
(0,0,-\frac{c}{(1-2 \epsilon)^2 (b-c)})\\
(0,\frac{2 c \epsilon}{(b-c)(1-2 \epsilon)},0)\\
(\frac{b(1-2\epsilon)-c}{2b(1-2\epsilon)},\frac{c}{b(1-2\epsilon)},0)\\
(\frac{b (1-2 \epsilon)-c}{(b-c)(1-2 \epsilon)},\frac{2 c \epsilon}{(b-c)(1-2\epsilon)},0)\\
(\frac{(b-c)(1-2\epsilon)^2-c}{(b-c)(1-2 \epsilon)^2},0,\frac{c}{(b-c)(1-2 \epsilon)^2})\\
(0,\frac{c \left[2(b-c) \epsilon (1-2 \epsilon)-c\right]}{(1-2 \epsilon) \left[(b-c)^2(1-2\epsilon) - bc\right]},\frac{c \left[b (2 \epsilon-1)+c\right]}{(1-2 \epsilon) \left[(b-c)^2(1-2\epsilon) + bc\right]})\\
(\frac{(b-c) \left[b (1-2 \epsilon)-c\right]}{(b-c)^2(1-2\epsilon)+bc},
\frac{c \left[2(b-c)\epsilon(1-2\epsilon)+c\right]}{(1-2 \epsilon)\left[(b-c)^2(1-2\epsilon)+bc\right]},\\
\frac{c \left[b(1-2\epsilon)-c\right]}{(1-2 \epsilon)
\left[(b-c)^2(1-2\epsilon)-bc\right]}).
\end{cases}.
\label{eq:fixedpoint}
\end{equation}

\begin{figure}
    \begin{tabular}{cc}
        \includegraphics[width=\columnwidth]{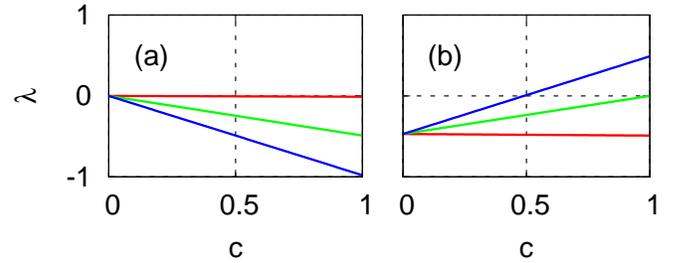}
    \end{tabular}
    \caption{(Color online)
        (a) Eigenvalues for the first FP of Eq.~\eqref{eq:fixedpoint},
        $\mathbf{x}_0^\ast = (x_{_{\rm AllD}}^\ast, x_{_{\rm TFT}}^\ast,
        x_{_{\rm WSLS}}^\ast) = (1,0,0)$. All of the eigenvalues
        are negative for the entire range of $c$. We set $b=1$ without loss
        of generality and choose $\epsilon=10^{-2}$.
        (b) Eigenvalues for another FP $(0,0,1)$. The largest
        eigenvalue is negative for $c < c_{\rm th}$. The threshold value
        $c_{\rm th}$ is analytically obtained in Eq.~\eqref{eq:c_th}. }
\label{fig:eig}
\end{figure}

By the linear stability analysis, we find that only one FP
$\mathbf{x}^*_0=(1,0,0)$, implying $x_{_{\rm AllD}}=1$, is stable for the
entire range of $c$, confirmed by the negativity of all eigenvalues, as shown in
Fig.~\ref{fig:eig}(a). This result is somehow consistent with the fact that this
FP is numerically observed for the entire range of $c$ even when
$\mu>0$, as depicted in Fig. 1(a) in the main text. We are also interested in
another FP $\mathbf{x}^*_0=(0,0,1)$, i.e., $x_{_{\rm WSLS}}=1$. It
is found that the largest eigenvalue is negative for $c<c_{\rm th}$ with some
threshold $c_{\rm th}$, implying that the FP is stable. However, it
becomes positive for $c>c_{\rm th}$, implying unstability of the FP.
See Fig.~\ref{fig:eig}(b). This largest eigenvalue is calculated as
\begin{equation}
    \lambda= \left(\epsilon-\frac{1}{2}\right) \left\{b (1-2 \epsilon)^2-2 c [2
    (\epsilon-1) \epsilon+1]\right\},
\end{equation}
from which we can define $c_{\rm th}$ as follows:
\begin{equation}
    c_{\rm th} = \frac{b(1-2 \epsilon)^2}{2(1-2 \epsilon +2 \epsilon^2)} \approx b\left(\frac{1}{2} - \epsilon\right).
    \label{eq:c_th}
\end{equation}
All the other FP's turn out to be unstable.

\section{Positive Lyapunov exponents}
\label{sect:lyap}

For chaotic dynamics, the largest Lyapunov exponent is positive.
The problem arises when its value is so small that the
numerical error of the algorithm has the same order of magnitude.
Suppose that we have a continuous orbit in a bounded three-dimensional region
and its Lyapunov exponents are sorted in descending order, i.e., $\lambda_1 >
\lambda_2 > \lambda_3$. It is well-known that one of the Lyapunov exponents must
be zero for a continuous orbit. We can also be sure that $\lambda_3$ must be
negative for the following reason: the information dimension of an attractor is
bounded from above by the Kaplan-Yorke formula,
\begin{equation}
D_{\text{KY}} = j + \frac{\sum_{i=1}^j \lambda_i}{|\lambda_{j+1}|},
\end{equation}
where $j$ satisfies $\sum_{i=1}^j \lambda_i > 0$ and $\sum_{i=1}^{j+1} \lambda_i
< 0$. In our case, the dimension cannot be greater than three, which means that
$j \le 2$. From $\sum_{i=1}^3 \lambda_i < 0$, it follows that $\lambda_3 < 0$.

Based on these results, we draw the following conclusions:
if we have a chaotic orbit, it implies that $\lambda_1 > 0$ whereas $\lambda_2$
is zero. Therefore, we have the following inequality,
\begin{equation}
\lambda_1 + \lambda_2 > 0.
\label{eq:ineq}
\end{equation}
On the other hand, if $\lambda_1$ is actually zero but numerically estimated as
a tiny positive number, $\lambda_1 + \lambda_2$ is likely to be negative because
of $\lambda_2$. The point is that Eq.~\eqref{eq:ineq} provides a sharp
criterion to check the positivity of $\lambda_1$, which can sometimes be obscure
if we observe $\lambda_1$ only. The Lyapunov exponents depicted in Fig. 1 of the
main text are estimated as positive in this way.

\section{Perturbative analysis for small positive $\mu$}
\label{sect:pert}

Based on the understanding of the mutation-free case in Appendix~\ref{sect:mu0},
we now analyze a more general case with $\mu>0$. As it is not straightforward to
obtain the FP's of the RD in Eq. (3) in the main text, we take the
perturbative approach. More specifically, we apply Newton's method in which each
FP $\mathbf{x}^*_0$ for $\mu = 0$ serves as a trial
solution. The first order Taylor expansion around the FP $\mathbf{x}_0^* =
(x_{_{0,\rm AllD}}^*,x_{_{0,\rm TFT}}^*,x_{_{0,\rm WSLS}}^*)$ yields
the following equation:
\begin{align}
&\mathbf{0} =
\begin{pmatrix}
f_{_{\rm AllD}}(\mathbf{x}^*)\\
f_{_{\rm TFT}}(\mathbf{x}^*)\\
f_{_{\rm WSLS}}(\mathbf{x}^*)
\end{pmatrix}\notag
=
\begin{pmatrix}
f_{_{\rm AllD}}(\mathbf{x}_0^*)\\
f_{_{\rm TFT}}(\mathbf{x}_0^*)\\
f_{_{\rm WSLS}}(\mathbf{x}_0^*)
\end{pmatrix}
\\& +
\begin{pmatrix}
\partial f_{_{\rm AllD}} \over \partial x_{_{\rm AllD}} && \partial f_{_{\rm AllD}} \over \partial x_{_{\rm TFT}} && \partial f_{_{\rm AllD}} \over \partial x_{_{\rm WSLS}} \\
\partial f_{_{\rm TFT}} \over \partial x_{_{\rm AllD}} && \partial f_{_{\rm TFT}} \over \partial x_{_{\rm TFT}} && \partial f_{_{\rm TFT}} \over \partial x_{_{\rm WSLS}} \\
\partial f_{_{\rm WSLS}} \over \partial x_{_{\rm AllD}} && \partial f_{_{\rm WSLS}} \over \partial x_{_{\rm TFT}} && \partial f_{_{\rm WSLS}} \over \partial x_{_{\rm WSLS}}
\end{pmatrix}
\begin{pmatrix}
x_{_{\rm AllD}}^* - x_{_{0,\rm AllD}}^*\\
x_{_{\rm TFT}}^* - x_{_{0,\rm TFT}}^*\\
x_{_{\rm WSLS}}^* - x_{_{0,\rm WSLS}}^*
\end{pmatrix},
\label{eq:newton}
\end{align}
where $\mathbf{x}^*=(x_{_{\rm AllD}}^*,x_{_{\rm TFT}}^*,x_{_{\rm WSLS}}^*)$ denotes the corresponding perturbative solution. The matrix in the above equation is indeed Jacobian matrix $\mathbf{J} = \left\{{\partial f_i \over \partial x_j } \right\}$ calculated at $\mathbf{x}=\mathbf{x}_0^*$. If $\mathbf{J}(\mathbf{x}_0^*)$ has an inverse, by arranging Eq.~(\ref{eq:newton}), we get
\begin{equation}
\begin{pmatrix}
x_{_{\rm AllD}}^*\\x_{_{\rm TFT}}^*\\x_{_{\rm WSLS}}^*
\end{pmatrix}
\approx
\begin{pmatrix}
x_{_{0,\rm AllD}}^*\\x_{_{0,\rm TFT}}^*\\x_{_{0,\rm WSLS}}^*
\end{pmatrix}
- \mathbf{J}^{-1}
\begin{pmatrix}
f_{_{\rm AllD}}(\mathbf{x}_0^*)\\
f_{_{\rm TFT}}(\mathbf{x}_0^*)\\
f_{_{\rm WSLS}}(\mathbf{x}_0^*)
\end{pmatrix}.
\label{eq:newton2b}
\end{equation}
Here, we would like to note that even when $\mathbf{J}(\mathbf{x}^*_0)$ has no
inverse, one can first calculate $\mathbf{J}(\mathbf{x})^{-1}$, then substitute
$\mathbf{x}$ by $\mathbf{x}^*_0$ to obtain the result $\mathbf{x}^*$ in
Eq.~(\ref{eq:newton2}b). Then, based on numerical observations, we pick up three
FP's for $\mu=0$, i.e., $(1,0,0)$, $(0,0,1)$, and
$(\frac{b(1-2\epsilon)-c}{2b(1-2\epsilon)},\frac{c}{b(1-2\epsilon)},0)$.
For example, the last one is chosen because it behaves similarly to
$\bar{\mathbf{x}}\approx (\frac{1-c}{2},c,\eta)$ with small positive $\eta$,
when the time average is taken over the cycle (see the main text).
We will denote this last FP as $\hat{\mathbf{x}}_0^\ast$.
By applying Newton's method to these FP's, we obtain the following
perturbative solutions:

\begin{widetext}
\begin{equation}
	\mathbf{x}^* \approx
	\begin{cases}

	\begin{pmatrix} 1\\0\\0\end{pmatrix} &+ \mu \begin{pmatrix}
			{\frac{(3 \epsilon+1) (b-c)}{3 c-6 c \epsilon}}\\
			{-\frac{(b-c)}{3 c-6 c \epsilon}}\\
			{\frac{2 \epsilon (b-c)}{3 c (2 \epsilon-1)}}
		\end{pmatrix}\\

	\begin{pmatrix}0\\0\\1\end{pmatrix} &+ \mu\begin{pmatrix}
			{-\frac{2  \left((\epsilon-1) \left(4 \epsilon^2-2 \epsilon+1\right) (b-c)\right)}{3 \left((2 \epsilon-1) \left(b (1-2 \epsilon)^2-2 c (2 (\epsilon-1) \epsilon+1)\right)\right)}}\\
			{\frac{1}{3} \left(\frac{1}{(2 \epsilon-1)^3}-1\right)}\\
			{\frac{2 (\epsilon-1) \left(4 \epsilon^2-2 \epsilon+1\right) \left(3 b^2 (1-2 \epsilon)^4-6 b c (1-2 \epsilon)^4+2 c^2 (12 (\epsilon-1) \epsilon (2 (\epsilon-1) \epsilon+1)+1)\right)}{3 (2 \epsilon-1)^3 \left(b (1-2 \epsilon)^2-4 c (\epsilon-1) \epsilon\right) \left(b (1-2 \epsilon)^2-2 c (2 (\epsilon-1) \epsilon+1)\right)}}
		\end{pmatrix}\\

	\begin{pmatrix}
	\frac{b(1-2\epsilon)-c}{2b(1-2\epsilon)}\\ \frac{c}{b(1-2\epsilon)}\\0\end{pmatrix} &+ \mu \begin{pmatrix}
			{\frac{(b-c) (2 \epsilon (b+c)-5 b-c) (b (2 \epsilon-1)+4 c)}{24 c^2 (2 \epsilon-1) (b (2 \epsilon-1)+c)}}\\
			{0}\\
			{-\frac{((b-c) (b+c) (b (2 \epsilon-1)+4 c))}{12 \left(c^2 (b (2 \epsilon-1)+c)\right)}}
			\end{pmatrix} + \begin{pmatrix}
               \frac{1}{8}\\0\\-\frac{1}{4}\end{pmatrix}.

	\end{cases}
\label{eq:sol_newton}
\end{equation}
\end{widetext}

We find that a term of $O(1)$ appears for $\hat{\mathbf{x}}_0^\ast$
because of $\mu \ll 1$, implying that the perturbative approach fails here.
The result can be understood by directly calculating
$\mathbf{J}(\mathbf{x}^*_0)$ for $\mu \rightarrow 0$:
\begin{equation}
\begin{aligned}
    &\mathbf{J}\left(\frac{b(1-2\epsilon)-c}{2b(1-2\epsilon)},\frac{c}{b(1-2\epsilon)},0\right)\\&=
{(c-b(1-2\epsilon))\over 4b^2}\begin{pmatrix}
2c^2 && c^2+b^2(1-2\epsilon) && c^2 \\
-4c^2 && -2c^2 && -2c^2\\
0 && 0 && 0
\end{pmatrix}.
\end{aligned}
\label{eq:jaco}
\end{equation}
This Jacobian matrix has zeros in the bottom row because $\nabla f_{_{\rm WSLS}}
\rightarrow 0$ in the limit of zero mutation.
In fact, one can readily check that $f_{_{\rm WSLS}}$ becomes a constant
function in this limit for every $\mathbf{x}$ such that
$x_{_{\rm TFT}} =1 -2 x_{_{\rm AllD}}$ and $x_{_{\rm WSLS}}=0$. The FP
$\hat{\mathbf{x}}_0^\ast$ also satisfies this condition, so every derivative
vanishes there along the direction of $(1, -2, 0)$. As a consequence,
even if we include higher-order derivatives in the Taylor expansion
[Eq.~\eqref{eq:newton}], the resulting matrices will always be singular. For
example, the Hessian matrix $\mathbf{H} \equiv \left\{{\partial^2 f_{_{WSLS}}
\over \partial x_i \partial x_j}\right\}$ for the second-derivative test
is obtained as
\begin{equation}
    \mathbf{H} (\hat{\mathbf{x}}_0^\ast) =
 \begin{pmatrix}
 0 && 0 && {c^2(-1+2\epsilon) \over b}  \\
 0 && 0 && {c^2(-1+2\epsilon) \over b} \\
 {c^2(-1+2\epsilon) \over b} && {c^2(-1+2\epsilon) \over 2b} && {c^2-b^2(1-2\epsilon)^2-bc(1-2\epsilon)^3 \over b}
 \end{pmatrix}
 \label{eq:hess}
\end{equation}
when $\mu \rightarrow 0$. It is straightforward to see that it has an
eigenvector $(1, -2, 0)$ associated with the eigenvalue zero, which makes the
test inconclusive.

\section{Symmetry of replicator dynamics}
\label{sect:symm}

Recall that RD in this work is defined as
\begin{equation}
    \frac{dx_i}{dt} = f_i \equiv (1-\mu) p_i x_i - \left< p \right> x_i +
    \frac{\mu}{|\mathcal{S}|-1} \sum_{j \neq i} p_j x_j,
\end{equation}
where $p_i = \sum_i a_{ij} x_j$ and $\left< p \right> \equiv \sum_i p_i x_i =
\sum_{ij} a_{ij} x_i x_j$. The long-term payoff $a_{ij}$ is given in
Table~\ref{ta:longtermpayoff}.
We explicitly write the functional dependence as $f_i(x_{_{\rm AllC}}, x_{_{\rm
AllD}}, x_{_{\rm WSLS}})$ by choosing three independent variables. Note that we
have just chosen $x_{_{\rm AllC}}$ as an independent variable instead of
$x_{_{\rm TFT}}$ as we have done in other parts of the paper. It is because this
set of independent variables show the symmetry most clearly. After some algebra,
we can readily see the following equality:
\begin{equation}
\begin{aligned}
f_{_{\rm AllC}} (x_1, x_2, &x_3) + f_{_{\rm AllD}} (x_2,x_1,x_3) \\ &= \frac{b-c}{3}
(1+x_3) [\mu (1+x_3)-x_1 (4\mu + 3x_3)].
\end{aligned}
\end{equation}
It is important to note that $x_1$ and $x_2$ exchange the positions in
evaluating $f_{_{\rm AllD}}$. Let us set $\mu$ and $x_3$ to zero so that the
right-hand side vanishes altogether. We may suppress the dependence on
$x_3$ which is fixed, so the result is
\begin{equation}
f_{_{\rm AllC}} (x_1, x_2) = - f_{_{\rm AllD}} (x_2,x_1).
\label{eq:zero}
\end{equation}
By construction, we already have
\begin{eqnarray}
\frac{dx_{_{\rm AllC}}}{dt} &=& f_{_{\rm AllC}} (x_{_{\rm AllC}}, x_{_{\rm
AllD}}) \label{eq:rd1}\\
\frac{dx_{_{\rm AllD}}}{dt} &=& f_{_{\rm AllD}} (x_{_{\rm AllC}}, x_{_{\rm
AllD}}).
\label{eq:rd2}
\end{eqnarray}
We combine Eqs.~\eqref{eq:zero} to \eqref{eq:rd2} to obtain
\begin{eqnarray}
\frac{dx_{_{\rm AllC}}}{dt} &=& -f_{_{\rm AllD}} (x_{_{\rm AllD}}, x_{_{\rm
AllC}})\\
\frac{dx_{_{\rm AllD}}}{dt} &=& -f_{_{\rm AllC}} (x_{_{\rm AllD}}, x_{_{\rm
AllC}}).
\end{eqnarray}
Now, let us define $\tau \equiv -t$ to rewrite the equations as
\begin{eqnarray}
\frac{dx_{_{\rm AllC}}}{d\tau} &=& f_{_{\rm AllD}} (x_{_{\rm AllD}}, x_{_{\rm
AllC}})\\
\frac{dx_{_{\rm AllD}}}{d\tau} &=& f_{_{\rm AllC}} (x_{_{\rm AllD}}, x_{_{\rm
AllC}}).
\end{eqnarray}
Finally, the variables are relabeled as $X_{_{\rm AlC}} \equiv x_{_{\rm AllD}}$
and $X_{_{\rm AlD}} \equiv x_{_{\rm AllC}}$, and we end up with the following
set of equations,
\begin{eqnarray}
\frac{dX_{_{\rm AllD}}}{d\tau} &=& f_{_{\rm AllD}} (X_{_{\rm AllC}}, X_{_{\rm
AllD}})\\
\frac{dX_{_{\rm AllC}}}{d\tau} &=& f_{_{\rm AllC}} (X_{_{\rm AllC}}, X_{_{\rm
AllD}}),
\end{eqnarray}
which are formally identical to the original ones [Eqs.~\eqref{eq:rd1} and
\eqref{eq:rd2}].

%
\end{document}